\documentclass[aps,prd,reprint,superscriptaddress,showpacs,amsmath,amssymb]{revtex4-2}

\usepackage[colorlinks,
           bookmarks=true,
           linkcolor=blue,
           urlcolor=blue, 
            anchorcolor=black,
            citecolor=blue
            ]{hyperref}

\usepackage{multirow}
\usepackage{graphicx}
\usepackage{dcolumn}
\usepackage{bm}
\usepackage{float}
\usepackage{exscale}
\usepackage{relsize}
\usepackage{amsmath}
\usepackage{amssymb}
\begin{document}

\title{Deciphering the nature of  $X(2300)$ with the PACIAE model}
\author{Jian Cao}
\affiliation{School of Physics and Information Technology, Shaanxi Normal University, Xi'an 710119, China}
\author{Wen-Chao Zhang}
\email{wenchao.zhang@snnu.edu.cn (corresponding author)}
\affiliation{School of Physics and Information Technology, Shaanxi Normal University, Xi'an 710119, China}
\author{Jin-Peng Zhang}
\affiliation{School of Physics and Information Technology, Shaanxi Normal University, Xi'an 710119, China}
\author{Bo Feng}
\affiliation{School of Physics and Information Technology, Shaanxi Normal University, Xi'an 710119, China}
\author{An-Ke Lei}
\affiliation{School of Physics and Electronic Science, Guizhou Normal University, Guiyang, 550025, China}
\author{Zhi-Lei She}     
\affiliation{Wuhan Textile University, Wuhan 430200, China}
\author{Hua Zheng}
\affiliation{School of Physics and Information Technology, Shaanxi Normal University, Xi'an 710119, China}
\author{Dai-Mei Zhou}
\email{zhoudm@mail.ccnu.edu.cn}
\affiliation{Key Laboratory of Quark and Lepton Physics (MOE) and Institute of
            Particle Physics, Central China Normal University, Wuhan 430079,
            China}
\author{Yu-Liang Yan}
\affiliation{China Institute of Atomic Energy, P. O. Box 275 (10), Beijing
            102413, China}
\author{Ben-Hao Sa}
\email{sabhliuym35@qq.com} 
\affiliation{China Institute of Atomic Energy, P. O. Box 275 (10), Beijing
            102413, China}   
\affiliation{Key Laboratory of Quark and Lepton Physics (MOE) and Institute of
            Particle Physics, Central China Normal University, Wuhan 430079,
            China}
\date{\today}

\begin{abstract}
Inspired by the BESIII newest observation of an axial-vector particle $X(2300)$ in the $\psi(3686)\rightarrow \phi\eta \eta'$ process [M. Ablikim et al. (BESIII Collaboration), Phys. Rev. Lett. 134, 191901 (2025)], we simulate its production in $e^+e^-$ collisions at $\sqrt{s}=4.95$ GeV using the parton and hadron cascade model PACIAE 4.0. In this model, the final partonic state (FPS) and the final hadronic state (FHS) are simulated and recorded sequentially. Beyond the existing assumptions of the $X(2300)$ as either an excited strangeonium state or an $ss\bar{s}\bar{s}$ tetraquark state, we propose, for the first time, that it could also be a $q\bar{q}s\bar{s}$ ($q=u/d$) tetraquark state or a hadro‑strangeonium state (a bound system of a strangeonium and a light hadron). The excited strangeonium candidate is formed by coalescing an $s\bar{s}$ quark pair in the FPS with the quantum statistical mechanics inspired dynamically constrained  phase-space coalescence (DCPC) model. The tetraquark candidates of $q\bar{q}s\bar{s}$ and  $ss\bar{s}\bar{s}$ are similarly produced by coalescing four constituent quarks in the FPS. In contrast, a hadro‑strangeonium candidate emerges from the recombination of the constituent mesons  $\phi$ and $\eta'/\eta$  in the FHS. We then calculate the $X(2300)$'s orbital angular momentum quantum number in its rest frame  and  perform the spectral classification for each of the above  candidates. Given its quantum numbers $J^{PC} = 1^{+-}$, the $X(2300)$  is identified as a $P$-wave $s\bar{s}$, an $S$-wave $q\bar{q}s\bar{s}/ss\bar{s}\bar{s}$ or an $S$-wave $\phi\eta'/\phi \eta$ candidate. For the first time, we estimate the production rates for these configurations. The $P$-wave $s\bar{s}$ and $S$-wave $q\bar{q}s\bar{s}$ states are produced at rates on the order of $10^{-5}$, whereas the $S$-wave $ss\bar{s}\bar{s}$ and $\phi\eta'/\phi \eta$ states appear at rates on the order of $10^{-6}$. Moreover, significant discrepancies are observed in the rapidity distributions and the transverse momentum spectra  among the different candidates. These discrepancies could be served as  valuable criteria for deciphering the nature of the $X(2300)$.
\end{abstract}

\maketitle

\section{Introduction}

In the constituent quark model \cite{quark_model_1, quark_model_2}, the conventional hadrons are classified into two broad families: mesons ($q\bar{q}$) and baryons ($qqq$ or $\bar{q}\bar{q}\bar{q}$). This model also permits the existence  of exotic hadrons such as tetraquarks ($qq\bar{q}\bar{q}$) and pentaquarks ($qqqq\bar{q}$). Exotic states provide a distinctive environment to investigate the strong interactions and the confinement mechanism \cite{exotic_meson}. The first exotic hadron $X(3872)$ was observed in $e^{+}e^{-}$ collisions by the Belle Collaboration in 2003 \cite{x3872_0}. Since then, many new exotic hadrons consistent with the tetraquark configuration have been observed \cite{pentaquark_1, pentaquark_2, pentaquark_3, pentaquark_4, pentaquark_5, pentaquark_6, pentaquark_7}. They include $X(2900)$ \cite{X2900}, $T_{cc}^+(3875)$ \cite{Tcc3900}, $Z_c(3900)$ \cite{Zc3900}, $X(6900)$ \cite{X6900_1, X6900_2, X6900_3}, etc. Among them, $X(6900)$ was interpreted as the fully charm tetraquark state $cc\bar{c}\bar{c}$ \cite{X6900_tetra_1, X6900_tetra_2, X6900_tetra_3, X6900_tetra_4,X6900_tetra_5}. Similar to the $cc\bar{c}\bar{c}$ state, it is reasonable to conjecture that the fully bottom tetraquark $bb\bar{b}\bar{b}$ and fully strange tetraquark $ss\bar{s}\bar{s}$ states might exist.


Recently, the BESIII collaboration reported the observation of an axial-vector particle $X(2300)$ in the $\phi\eta$ and $\phi \eta'$ invariant mass spectra for the $\psi(3686)\rightarrow \phi\eta \eta'$ process \cite{x_2300_BESIII}. Its mass and width are, respectively,  2316 MeV/c$^2$ and 89 MeV. Its spin and parity are determined as $J^{PC}=1^{+-}$. There are several theoretical speculations for the nature of the $X(2300)$, including the excited strangeonium state \cite{X2300_expla_1,X2300_expla_2, X2300_expla_3}, the tetraquark state ($ss\bar{s}\bar{s}$) \cite{X2300_expla_4,X2300_expla_5}, etc.  Clearly, distinguishing between the various models and deciphering the nature of the $X(2300)$ will require further theoretical and experimental efforts.

In this work, we simulate the production of the $X(2300)$ in $e^+e^-$ collisions at center-of-mass energy $\sqrt{s}=$ 4.95 GeV with a parton and hadron cascade model PACIAE 4.0 \cite{paciae_4}. In this model, the final partonic state (FPS) and the final hadronic state (FHS) are simulated and recorded sequentially. While previous interpretations have described the $X(2300)$ as an excited strangeonium state or a tetraquark $ss\bar{s}\bar{s}$ state \cite{X2300_expla_1,X2300_expla_2, X2300_expla_3, X2300_expla_4,X2300_expla_5}, we note that the $U(1)$ anomaly coupling permits non‑strange quarks to couple to a vector $s\bar{s}$ component via soft‑gluon interactions. This motivates us to propose, for the first time, that the $X(2300)$ could also be a $q\bar{q}s\bar{s}$ ($q=u/d$) state. Furthermore, we suggest another novel possibility that the $X(2300)$ may be a hadro‑strangeonium state, i.e., a bound system of a strangeonium and a light hadron. This state is similar to the hadro-charmonium state, whose attraction potential between a charmonium  and a light hadron is about a few hundred MeV \cite{hadro-charm_1, hadro-charm_2, hadro-charm_3}. In Refs. \cite{hadro-charm_2, hadro-charm_3}, the exotic hadron $P_c(4450)$ \cite{Pc_4450} was interpreted as a hadro-charmonium bound state of a charmonium and a nucleon. The excited strangeonium and tetraquark state of the $X(2300)$ candidates are, respectively, generated by coalescing with two quarks $s\bar{s}$ and four quarks of $q\bar{q}s\bar{s}/ss\bar{s}\bar{s}$ in the FPS using the quantum statistical mechanics inspired dynamically constrained phase-space coalescence (DCPC) model \cite{DCPC}. The hadro-strangeonium state of the $X(2300)$ candidate is produced by the recombination of the constituent mesons $\phi$ and $\eta'/\eta$   in the FHS. We then evaluate the orbital angular momentum  quantum number of the $X(2300)$ candidates in their rest frame and spectrally classify them according to the standard $n^{2S+1}L_J$ spectroscopic notation \cite{book1}, where $n$, $S$, $L$ and $J$ are, respectively, the radial excitation, the spin, the orbital angular momentum, and the total angular momentum quantum numbers of the $X(2300)$ candidates. The wave shapes of  $S$, $P$, $D$, ... corresponds to the $L=$ 0, 1, 2, ...,  respectively.

Given its $J^{PC} = 1^{+-}$ \cite{x_2300_BESIII}, the $X(2300)$ can be identified as either a $P$-wave $s\bar{s}$ strangeonium state, an $S$-wave $q\bar{q}s\bar{s}/ss\bar{s}\bar{s}$ tetraquark state, or an $S$-wave $\phi\eta'/\phi \eta$ hadro-strangeonium state. We will compare the production yields, rapidity distributions, and transverse momentum spectra of these different configurations. The observed differences are expected to provide valuable criteria for deciphering the nature of the $X(2300)$. 

\section{The model and methodology} \label{sec:method}

In the PACIAE 4.0 model, the simulation of $e^+e^-$ annihilation is first executed by PYTHIA8 \cite{pythia_8} with the presetting of the hadronization turning-off and the postsetting of the breaking-up of the strings and diquarks/anti-diquarks into quarks/anti-quarks.  After the execution, one  obtains a state composed of the quarks, anti-quarks, and the gluons. A process of the gluon breaking-up and energetic quark (antiquark)  deexcitation is then performed.  The FPS is generated after the partonic $2\rightarrow 2$ rescattering, where the leading order  perturbative Quantum Chromodynamics (QCD) parton-parton scattering cross sections \cite{cs_1,cs_2} are employed. It comprises numerous quarks and anti-quarks with their four-dimensional coordinates and momenta. An intermediate hadronic is produced with the hadronization of the FPS by the Lund string fragmentation scheme \cite{pythia_6}. Finally, this intermediate  state undergoes $2\rightarrow 2$ hadronic rescattering \cite{book2, paciae_3}, leading to kinetic freeze-out and producing the FHS. It is composed of abundant hadrons with their four-dimensional coordinate and momentum.

The DCPC model was proposed by us for the first time to study the light nuclei production in proton-proton collisions at the LHC energies \cite{DCPC}. Based on PACIAE generated final partonic or hadronic states, the DCPC model has been successfully employed to calculate the production yields of various exotic hadronic states. They include $X(3872)$ \cite{ge2021, tai2023, she2024}, $Z_c^{\pm}(3900)$ \cite{zc_3900}, $G(3900)$ \cite{G_3900}, $P_c(4312)$, $P_c(4440)$, $P_c(4457)$ \cite{hui2022},   $\Omega_c^0$ \cite{Omega_c}, and $X$(2370) \cite{zhang2024}. In this model, inspired by the the quantum statistical mechanics \cite{kobo1965, stowe2007}, the yield of $N$-particle cluster is estimated as
\begin{eqnarray}
Y_{N}=\int\cdots\int_{E_\alpha\le E\le E_\beta}\frac{d\boldsymbol{x}_{1}d\boldsymbol{p}_{1}\cdots d\boldsymbol{x}_{N}d\boldsymbol{p}_{N}}{h^{3N}},
\label{eq: two}
\end{eqnarray}
where $E_{\alpha}$ and $E_{\beta }$ denote the lower and upper
energy thresholds of the cluster, respectively. $\boldsymbol{x}_{i}$ and $\boldsymbol {p}_{i}$ are, respectively, the three-dimensional coordinate and momentum of the $i$-th particle in the centre-of-mass system (cms) of the $e^+e^-$ collision.  It is postulated that a naturally formed cluster must inherently satisfy specific dynamical constraints, namely those pertaining to the components, the component coordinates, and the component momenta. For instance, the yield of the $X(2300)$ hadro-strangeonium state composed of the bound $\phi\eta'$ or $\phi\eta$ reads
\begin{eqnarray}
Y_{\phi\eta'/\phi\eta}=\int\cdots\int\delta_{12}\frac{d\boldsymbol{x}_{1}d\boldsymbol{p}_{1} d\boldsymbol{x}_{2}d\boldsymbol{p}_{2}}{h^{6}}.
\label{eq: two}
\end{eqnarray}
Here the dynamical constraint $\delta_{12}$ is expressed as

\begin{equation}
\delta_{12}=
\begin{cases}
1,& \textrm {if} \ 1\equiv \phi,\ 2\equiv  \eta'\ \textrm{or}\ \eta,\ R_i\leq  R_0,\ \textrm {and} \atop  m_{\rm 0}-\Delta m \leq m_{\rm inv}\leq m_{\rm 0}+\Delta m   \\ 
0,&  \ \ \ \ \ \ \ \ \ \rm otherwise
\end{cases},\label{eq:delta}
 \end{equation}
where $m_0$ is the mass of the $X(2300)$ candidate, $\Delta m$ refers to the mass
uncertainty (a free parameter) which is estimated as the half decay width of the $X(2300)$ candidate.
$R_0$ denotes the radius of the cluster, which is a free parameter. $R_i$ ($i=1, 2$) is defined as the magnitude of the position vector for the $i$-th constituent meson, measured in the rest frame of the cluster. In this frame, the three-dimensional coordinate of the constituent meson ($\phi\eta'$ or $\phi\eta$) $\boldsymbol{x}^{*}_i$ is obtained by first Lorentz transforming $\boldsymbol{x}_i$ to the cms of constituent mesons and then propagating the earlier freeze-out component meson freely to the freeze-out time of the later component meson \cite{zhull1,zhull2}. $R_0$ is assumed to be in the range of 1 fm $<R_0<$ 2 fm, where the lower (upper) limit is taken as the radius of one meson (the radius summation of two mesons). The invariant mass ($m_{\rm inv}$) is evaluated as
\begin{eqnarray}
m_{\textrm{inv}}=\sqrt{\bigg(\sum^{2}_{i=1} E_i \bigg)^2-\bigg(\sum^{2}_{i=1}
\boldsymbol{p}_i \bigg)^2},
\label{eq:two_3}
\end{eqnarray}
where $E_i$ and $\boldsymbol{p}_i$ ($i=$ 1, 2) are the energy and the three-dimensional momentum of the component meson ($\phi\eta'$ or $\phi\eta$)  in the cms of the $e^+e^-$ collisions, respectively. The yield of the $X(2300)$ excited strangeonium or tetraquark state can be evaluated in a similar way with different parameters. The parameters of mass uncertainty and radius are given in Table \ref{tab:X_2300_para}.

\begin{table}[h]
\caption{The parameters of the mass uncertainty and the radius for the $X(2300)$ candidates of the
 excited $s\bar{s}$ strangeonium state, the $q\bar{q}s\bar{s}/ss\bar{s}\bar{s}$ tetraquark state, and the  $\phi\eta$ and $\phi\eta'$ hadro-strangeonium state.}\label{tab:X_2300_para} 
\begin{ruledtabular}
\begin{tabular}{cccccc}
                  &     excited    $s\bar{s}$      &  $q\bar{q}s\bar{s}$ &  $ss\bar{s}\bar{s}$  & $\phi\eta'$         & $\phi\eta$        \\
 \colrule                 
 $\Delta m$               &  \multirow{2}{*}{44.5} & \multirow{2}{*}{44.5}                   & \multirow{2}{*}{44.5}                  & \multirow{2}{*}{44.5}          & \multirow{2}{*}{44.5}          \\
 (MeV$/c^2$)                 &                    &                   &          &           \\
 \hline
$R_0$ (fm)                & 1.0            &  1.0    & 1.0                  & 1.0-2.0         & 1.0-2.0        
\end{tabular}
\end{ruledtabular}
\end{table}

To generate the $X(2300)$ hadro-strangeonium state for instance, the following procedure is adopted. First, a list of component mesons, including $\phi$, $\eta'$, and $\eta$, is constructed using the FHS simulated by the PACIAE model. A double-loop iteration is then performed over all mesons in this list. Each combination, if it corresponds to either $\phi\eta'$ or $\phi\eta$ and satisfies the constraints given in Eq. (\ref{eq:delta}), then is accepted as an $X(2300)$ candidate. The constituent mesons of each accepted candidate are subsequently removed from the list. The double-loop process is repeated on the updated list until no mesons remain or no further valid candidates can be formed. The production of the $X(2300)$ excited strangeonium or tetraquark state is done in a similar way. The couplings for both the production and decay of the $X(2300)$  are not provided, as they are already embedded through the geometric constraints, kinematic constraints, and coalescence mechanisms in the DCPC model. 

To spectrally classify the $X(2300)$ candidate, its orbital angular momentum quantum number must be determined. The details of the method can be found in our previous work \cite{G_3900}. Here we describe it briefly. In the rest frame of the $X(2300)$ candidate, its orbital angular momentum (OAM) $\boldsymbol{l}^{*}$ is defined as the sum of the OAMs of its constituents:
\begin{eqnarray}
\boldsymbol{l}^{*}=\sum_{i=1}^{N}\boldsymbol{x}_i^{*}\times \boldsymbol{p}^{*}_i,
\label{eq:OAM}
\end{eqnarray}
where $N$ is the number of constituents in a given configuration of the $X(2300)$ candidate, $\boldsymbol{p}_i^{*}$ ($i = 1, 2, \cdots, N$) denotes the three-dimensional momentum of the $i$-th constituent in this frame. It is obtained by applying a Lorentz transformation of $\boldsymbol{p}_i$ into the cms frame of the constituents (i.e., the $X(2300)$ candidate rest frame). According to quantum mechanics, the orbital angular momentum is quantized, satisfying the relation:
\begin{equation}
\boldsymbol{l}^{*2} = L(L+1)\hbar^2,
\label{eq:OAM_number}
\end{equation}
where $\hbar$ is the reduced Planck constant and $L$ denotes the orbital angular momentum quantum number of the $X(2300)$ candidate. Since $L$ must be an integer, Eq. (\ref{eq:OAM_number}) is solved as
\begin{equation}
L = \mathrm{round}\left( \frac{-1 + \sqrt{1 + 4\boldsymbol{l}^{*2} / \hbar^{2}}}{2} \right),
\label{eq:OAM_method}
\end{equation}
where the function $\mathrm{round}(X)$ returns the integer closest to $X$.

For the $X(2300)$ interpreted as a hadro-strangeonium state ($\phi\eta'$ or $\phi\eta$ bound system), its parity is given by $P = P_\phi \cdot P_{\eta'/\eta} \cdot (-1)^L =(-1)^L$. If the $X(2300)$ is instead considered as an excited $s\bar{s}$ strangeonium state, its parity becomes $P = P_s \cdot P_{\bar{s}} \cdot (-1)^L =(-1)^{L+1}$. In the tetraquark picture ($q\bar{q}s\bar{s}/ss\bar{s}\bar{s}$), the parity is $P = P_{q/s} \cdot P_{\bar{q}/\bar{s}}\cdot  P_s  \cdot P_{\bar{s}} \cdot (-1)^L = (-1)^L$. For the  $s\bar{s}$, $q\bar{q}s\bar{s}$, and $ss\bar{s}\bar{s}$ configurations, the $C$-parity follows the common form: $C = (-1)^{L + S}$, where $S$ denotes the total spin of the constituent particles. For the $\phi\eta'$ or $\phi\eta$ system, the $C$-parity is expressed as $C = C_{\phi}\cdot C_{\eta'/\eta}\cdot(-1)^{L}=(-1)^{L+1}$. For Given the spin $S$ and orbital angular momentum quantum number $L$ of an $X(2300)$ candidate, its total angular momentum $J$ can take on the values $J = |L - S|, |L - S| + 1, \dots, L + S$.

\begin{table}[]
\caption{The $J^{PC}$s for the $S$-, $P$-, and $D$-wave $X(2300)$ candidates with different configurations.}\label{tab:X_2300_JPC}
\begin{ruledtabular}
\begin{tabular}{ccccc}
                  &     & $S$-wave & $P$-wave & $D$-wave \\ \hline
\multirow{2}{*}{$s\bar s$} & $S=0$ & $0^{-+}$  & $1^{+-}$  & $2^{-+}$   \\
                  & $S=1$ &  $1^{--}$ & $(0,1,2)^{++}$  &  $(1,2,3)^{--}$ \\  \hline
\multirow{3}{*}{$q\bar{q}s\bar{s}/ss\bar{s}\bar{s}$} & $S=0$ &  $0^{++}$ & $1^{--}$  &  $2^{++}$ \\
                  & $S=1$ & $1^{+-}$  & $(0,1,2)^{-+}$  &  $(1,2,3)^{+-}$ \\
                  & $S=2$ & $2^{++}$  & $(1,2,3)^{--}$  &  $(0,1,2,3,4)^{++}$ \\  \hline
                $\phi\eta'/\phi\eta$  & $S=1$ & $1^{+-}$  & $(0,1,2)^{-+}$  & $(1,2,3)^{+-}$ 
\end{tabular}
\end{ruledtabular}
\end{table}

\begin{table*}[]
\caption{The event-average yields of the $P$-wave $s\bar s$ strangeonium state, the $S$-wave $q\bar{q}s\bar{s}/ss\bar s\bar s$ tetraquark state, and the $S$-wave $\phi\eta'/\phi\eta$ hadro-strangeonium  state   for the $X(2300)$ candidates in $e^+e^-$ collisions at $\sqrt{s} = 4.95\text{ GeV}$. The uncertainties quoted are statistical errors. }\label{tab:X_2300_yield}
\begin{ruledtabular}
\begin{tabular}{cccccc}
  & $P$-wave  & \multicolumn{2}{c}{$S$-wave}  & \multicolumn{2}{c}{$S$-wave} \\\cline{3-4} \cline{5-6}
 & $s\bar s$ & $q\bar{q}s\bar{s}$ &$ss\bar  s\bar s$ &       $\phi\eta'$      &  $\phi\eta$        \\\hline
yield & $(4.05\pm0.01)\times 10^{-5}$ & $(8.43\pm0.02)\times 10^{-5}$ & $(6.50\pm0.06)\times 10^{-6}$         & $(6.97\pm0.06)\times 10^{-6} $  &$(6.06\pm 0.06)\times 10^{-6} $     
\end{tabular}
\end{ruledtabular}
\end{table*}

\section{Results and discussions} 
The $e^+e^-$ collisions at $\sqrt{s} = 4.95\text{ GeV}$ are simulated using the PACIAE 4.0 model \cite{paciae_4}. A total of 2 billion events are generated with the default model parameters. The excited strangeonium and tetraquark state of the $X(2300)$ candidates are, respectively, produced by coalescing with two quarks $s\bar{s}$ and four quarks $q\bar{q}s\bar{s}/ss\bar{s}\bar{s}$ in the FPS with the DCPC model. The hadro-strangeonium state of the $X(2300)$ candidate is generated by the recombination of the component mesons $\phi$ and $\eta'/\eta$ in the FHS with DCPC. Table \ref{tab:X_2300_JPC} presents the $J^{PC}$s for the $S$-, $P$-, and $D$-wave $X(2300)$ candidates with different configurations. They are determined according to the method in section \ref{sec:method}. Given its quantum numbers $J^{PC}=1^{+-}$ \cite{x_2300_BESIII}, the $X(2300)$ may be interpreted as a $P$-wave $s\bar{s}$ state, an $S$-wave $q\bar{q}s\bar{s}/ss\bar{s}\bar{s}$, or an $S$-wave $\phi\eta'/\phi \eta$. Within either the tetraquark or hadro-strangeonium picture, a $D$-wave configuration could also yield $J^{PC}=1^{+-}$. However, for the tetraquark interpretation, the nonrelativistic potential quark model predicts a mass of approximately 2600 MeV$/c^2$ for such a $D$-wave $ss\bar s\bar s$ state \cite{X2300_expla_4}, which is significantly higher than the observed mass of $X(2300)$. Moreover, forming a $D$-wave hadro-strangeonium state is considerably more difficult compared to its $S$-wave counterpart. As a result, neither the $D$-wave tetraquark nor the $D$-wave hadro-strangeonium hypothesis will be considered further in this work.

According to the BESIII data presented in Ref. \cite{x_2300_BESIII}, the production rates of $\phi(2170)$ (also denoted $Y(2175)$) in the $\phi\eta'$ and $\phi\eta$ channels are comparable to those of the $X(2300)$. The spin‑parity of $\phi(2170)$ has been determined as $J^{PC}=1^{--}$ \cite{phi_2170}. In Refs. \cite{phi_2170_exp1, phi_2170_exp2}, $\phi(2170)$ has been interpreted as an excited $s\bar{s}$ state, specifically the $2^3D_1$ or $3^3S_1$ configuration. For an $s\bar{s}$ system, the parity is given by $P = (-1)^{L+1}$. Thus, to obtain the negative parity of $\phi(2170)$, $L$ must be an even number, such as 0 or 2, which corresponds precisely to an $S$- or $D$-wave assignment as discussed in those references. In the case of the $X(2300)$, we have assumed it could be a $P$-wave $s\bar{s}$ state. This picture aligns with the interpretation in Ref. \cite{X2300_expla_2}, where the authors identify the $X(2300)$ as the $3^1P_1$ state. The mass of a $3^1 P_1$  $s\bar{s}$ configuration is expected to be larger than that of a $2^3 D_1$ or $3^3 S_1$ state, which is consistent with the observed mass ordering $m(\phi(2170)) < m(X(2300))$. Therefore, the $X(2300)$ can naturally be regarded as an excited $P$-wave $s\bar{s}$ state. Based on quantum-number considerations, the $X(2300)$ can be regarded as a candidate for a $\phi\eta'/\phi \eta$ hadro‑strangeonium state, while the $\phi(2170)$ cannot. The key distinction lies in the $P$ and $C$ quantum numbers of a $\phi\eta'/\phi \eta$ system. For such a system, the parity is $P = (-1)^{L}$ and the charge‑conjugation parity is $C=(-1)^{L+1}$. Consequently, for any $L$, the resulting $P$ and $C$ are necessarily opposite in sign. Given the established quantum numbers $1^{+-}$ for $X(2300)$ and $1^{--}$ for $\phi(2170)$, only the $X(2300)$ is compatible with the $P$-$C$ relation of the $\phi\eta'/\phi \eta$ system.

The event-average yields of the $P$-wave $s\bar s$ strangeonium state, the $S$-wave $q\bar{q}s\bar{s}/ss\bar{s}\bar{s}$ tetraquark state, and the $S$-wave $\phi\eta'/\phi\eta$ hadro-strangeonium  states   for the $X(2300)$ candidates are estimated for the first time and summarized in  Table \ref{tab:X_2300_yield}. The yield quoted for $q\bar{q}s\bar{s}$  corresponds to the sum of the yields of $u\bar{u}s\bar{s}$ and $d\bar{d}s\bar{s}$. We observe that the yields of the $P$-wave $s\bar{s}$ and $S$-wave $q\bar{q}s\bar{s}$ states are on the order of $10^{-5}$, whereas those of the $S$-wave $ss\bar{s}\bar{s}$ and $\phi\eta'/\phi \eta$ states are on the order of $10^{-6}$. Notably, the yield of the $S$-wave $ss\bar{s}\bar{s}$ state is approximately twice that of the $P$-wave
$s\bar{s}$ state. Furthermore, the $\phi\eta$ state shows a higher yield than $\phi\eta'$. This is consistent with the fact that $\eta'$ has a larger mass than $\eta$, resulting in a more suppressed production rate due to phase space limitations.

We have studied the influence of the coalescence radius $R_0$ on the production yield of $X(2300)$ candidates in various configurations. For the hadro-strangeonium state, raising the lower bound of $R_0$ from 0.7 fm to 1.0 fm reduces the yield of $\phi\eta'$ from $1.26\times 10^{-5}$ to $6.06\times 10^{-6}$, and that of $\phi\eta$ from $1.42\times 10^{-5}$ to $6.97\times 10^{-6}$. Conversely, increasing the upper bound of $R_0$ from 2 fm to 2.3 fm raises the yield of $\phi\eta'$ from $6.06\times 10^{-6}$ to $6.43\times 10^{-6}$, and that of $\phi\eta$ from $6.97\times 10^{-6}$ to $7.43\times 10^{-6}$. For the tetraquark and excited strangeonium states, increasing the upper bound of $R_0$ from 0.7 fm to 1.0 fm enhances the yields of $q\bar{q}s\bar{s}$ and $ss\bar{s}\bar{s}$ from $1.52\times 10^{-6}$ and $1.45\times 10^{-7}$  to $8.43\times 10^{-5}$ and $6.50\times 10^{-6}$, respectively, while the yield of the excited $s\bar s$ state rises from $2.30\times 10^{-6}$ to $4.05\times 10^{-5}$. Furthermore, to examine the impact of different methods to extract the value of $L$  for the $X(2300)$ candidates, we introduced an alternative function “trunc($X$)” in Eq. (\ref{eq:OAM_method}), which truncates $X$ to an integer by discarding its fractional part. Switching from  “round($X$)” to “trunc($X$)” increases the yields of the $s\bar s $, $q\bar{q}s\bar{s}$, $ss\bar{s}\bar{s}$, $\phi\eta'$, and $\phi\eta$  by approximately a factor of 1.08, 2.80, 2.34, 2.11, and 2.25 respectively.

\begin{figure}[]
\centering
\includegraphics[scale=0.461]{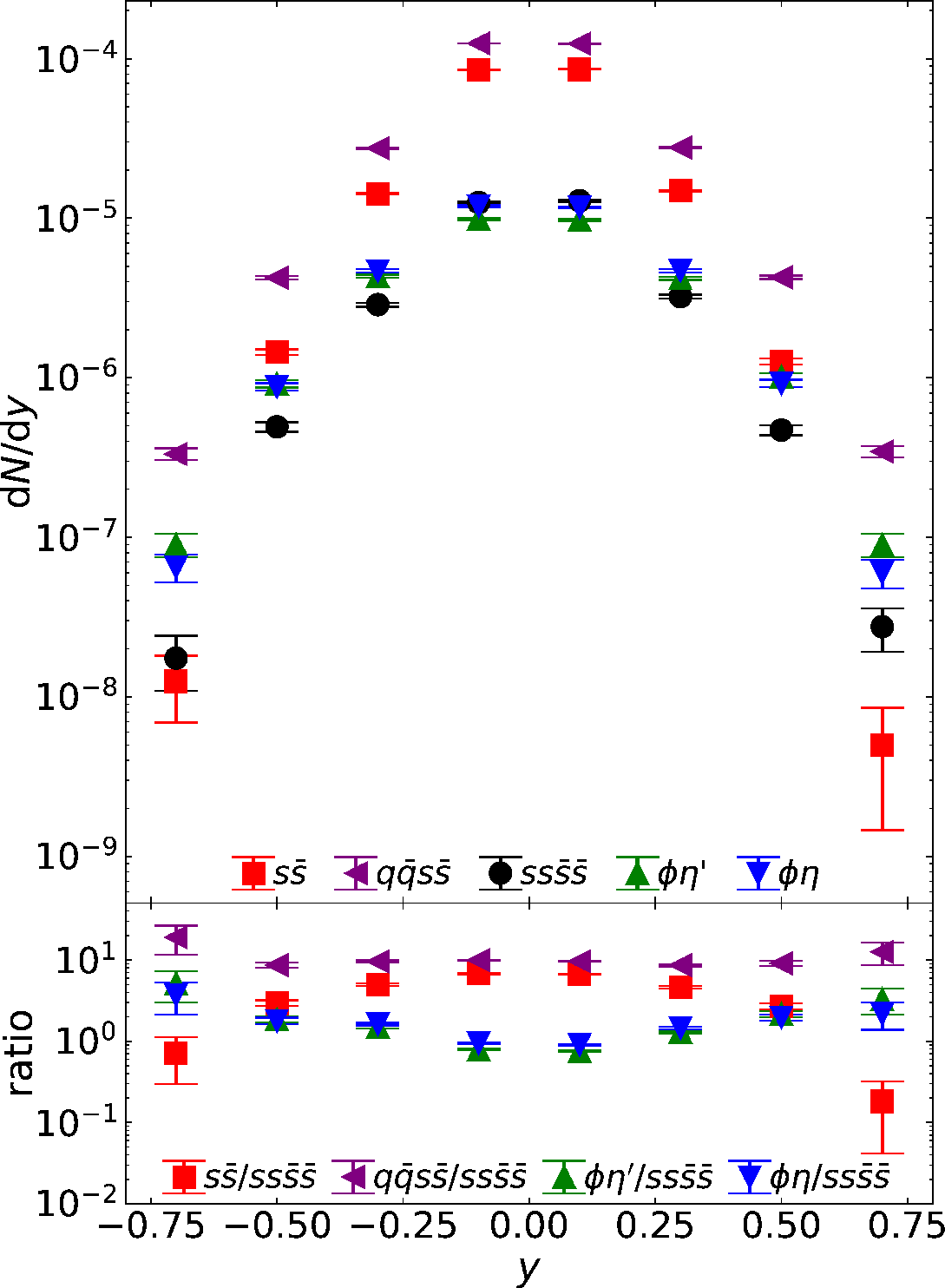}
\caption{\label{fig:yy_X2300}  Upper panel: the simulated $y$ 
single-differential distributions for the $P$-wave $s\bar{s}$ (squares), the $S$-wave $q\bar{q}s\bar{s}$ (left triangles) and $ss\bar{s}\bar{s}$ (circles), as well as the $S$-wave $\phi\eta'$ (upward triangles) and $\phi\eta$ (downward triangles) states of the $X(2300)$ candidates in $e^+e^-$ collisions at $\sqrt{s}$= 4.95 GeV. Lower panel: the ratios between two distributions denoted by legend. The error bars represent the statistical uncertainties.}
\end{figure}

\begin{figure}[H]
\centering
\includegraphics[scale=0.461]{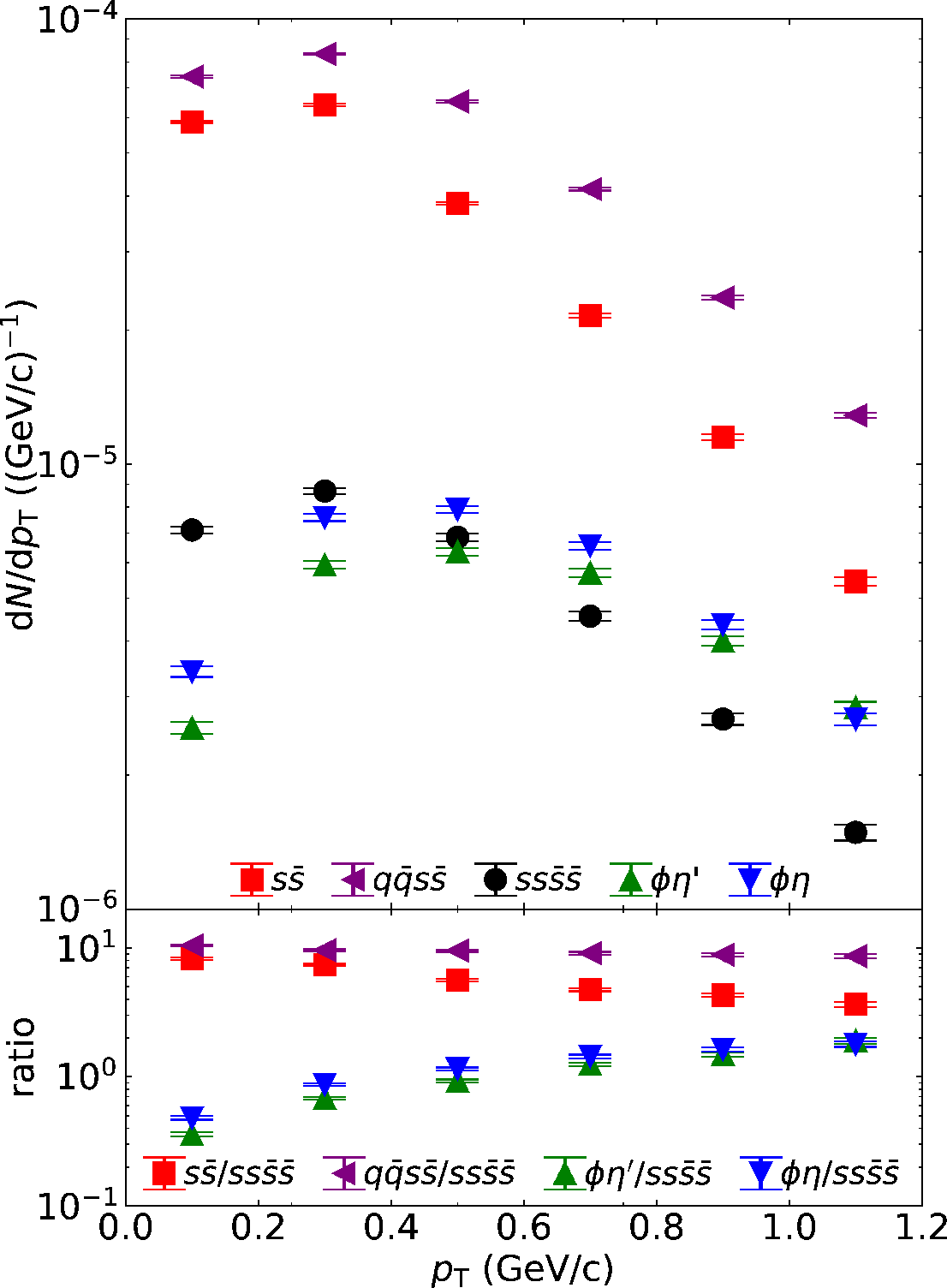}
\caption{\label{fig:pt_X2300} Similar as that in Fig. \ref{fig:yy_X2300}, but for the $p_{\rm T}$-differential cross sections of  the $X(2300)$ candidates with different configurations in $e^+e^-$ collisions at $\sqrt{s}$= 4.95 GeV.}
\end{figure}

The upper panels in Figs. \ref{fig:yy_X2300} and  \ref{fig:pt_X2300} show, respectively, the simulated $y$  and $p_{\rm T}$ single-differential distributions for the $P$-wave $s\bar{s}$ (squares), the $S$-wave $q\bar{q}s\bar{s}$ (left triangles) and $ss\bar{s}\bar{s}$ (circles), as well as the $S$-wave $\phi\eta'$ (upward triangles) and $\phi\eta$ (downward triangles) states of the $X(2300)$ candidates in $e^+e^-$ collisions at $\sqrt{s}$= 4.95 GeV. In the upper panels of Fig. \ref{fig:yy_X2300}, there is a peak at mid-rapidity for the rapidity distributions of the $X(2300)$ candidates with different configurations. This could be understood as follows. The central rapidity region  accumulates particles from all event types, regardless of jet direction. In contrast, the large-rapidity region receives contributions only from events with the jets aligned close to the beam axis. This leads to a significantly higher particle yield at central rapidity, resulting in a distinct peak. Furthermore, the peak height follows the hierarchy $q\bar{q}s\bar{s}>s\bar{s} > ss\bar{s}\bar{s} > \phi\eta > \phi\eta'$, which is consistent with  the hierarchy of their production yields. In the upper panel of Fig. \ref{fig:pt_X2300}, there is a peak at low $p_{\rm T}$ for the $p_{\rm T}$ single-differential distributions of the $X(2300)$ candidates with different configurations. It is a consequence of the non-perturbative nature of soft QCD and the process of hadronization. Moreover, the peak in the $p_{\rm T}$ distribution for the $\phi\eta'/\phi\eta$ configuration is observed at approximately 0.5 GeV/c, which is notably higher than that of the compact $q\bar{q}s\bar{s}/ss\bar s\bar s/s\bar s$ state, peaking at around 0.3 GeV/c. This difference is consistent with theoretical expectations based on their distinct internal structures and production mechanisms. $\phi$ and $\eta'/\eta$ are both heavy mesons. Their co-production as a hadro-strangeonium state requires substantial transverse momentum transfer from the parton shower. As a result, the resulting $X(2300)$ candidate retains this collective transverse boost, shifting its $p_{\rm T}$ peak to a higher value. In contrast, the $q\bar{q}s\bar{s}/ss\bar s\bar s/s\bar s$ state is assembled directly via the quark recombination in the FPS. Its production does not involve the independent formation and kinematic coordination of two pre-existing heavy mesons. It acquires the transverse momentum from the global kick of the parton shower. Although its mass is similar to the hadro-strangeonium state, the absence of the additional co-production constraint prevents their $p_{\rm T}$ spectrum from being biased toward higher values. Hence, their peak naturally lies in the kinematic region associated with soft processes, which corresponds to lower transverse momentum.

The obvious discrepancies among the yields, the rapidity distributions,  and the $p_{\rm T}$ spectra for the $X(2300)$ candidates with different configurations provide crucial  criteria for deciphering the nature of the $X(2300)$. We strongly suggest the experimental measurement of these variables of the $X(2300)$ candidates in  the $e^+e^-$ annihilations at the BESIII energies and comparing them with our  results.

\vspace{3mm}

\section{acknowledgments}
We would like to thank Prof. Xian-Hui Zhong at Hunan Normal University for the fruitful discussions.  This work is supported by the National Natural Science
Foundation of China under grant Nos. 11447024, 11505108 and 12375135, and by the 111 project of the
foreign expert bureau of China. Y.L.Y. acknowledges the financial support
from Key Laboratory of Quark and Lepton Physics in Central
China Normal University under grant No. QLPL201805 and the Continuous Basic
Scientific Research Project (No, WDJC-2019-13). W.C.Z. is supported
by the Natural Science Basic Research Plan in Shaanxi Province of China
(No. 2023-JCYB-012). H.Z. acknowledges the financial support from
Key Laboratory of Quark and Lepton Physics in Central China Normal University
under grant No. QLPL2024P01.


\end{document}